\documentclass[aps,prl,longbibliography,twocolumn,superscriptaddress,amsfont,graphicx,nofootinbib,preprintnumbers]{revtex4-1}%
\UseRawInputEncoding
\usepackage{color,graphicx,epsfig}
\usepackage{ifpdf}
\usepackage{amsmath}
\usepackage{bm}
\usepackage[english]{babel}
\usepackage{amssymb}
\usepackage{braket}
\usepackage{setspace}
\allowdisplaybreaks[4]

\usepackage{hyperref}
\usepackage{enumerate}
\usepackage{lipsum}

\usepackage{svg}
\svgsetup{
    inkscapepath=i/svg-inkscape/
}
\svgpath{{svg/}}
\usepackage{xcolor}
\usepackage{setspace}
\usepackage{hyperref}

\usepackage[utf8]{inputenc}
\usepackage[T1]{fontenc}
\usepackage{lmodern}

\usepackage{booktabs}
\usepackage{color}
\usepackage{epsfig}
\usepackage{ifpdf}
\usepackage{amsmath}
\usepackage{bm}
\usepackage[english]{babel}
\usepackage{amsfonts}
\usepackage{amssymb}
\usepackage{braket}
\usepackage{enumerate}
\usepackage{cancel}
\usepackage{multirow}
\usepackage{cleveref}
\usepackage{xspace}
\usepackage{array}
\usepackage{fancyvrb}
\usepackage{fontawesome}
\usepackage{dcolumn}
\usepackage{slashed}
\usepackage[normalem]{ulem}
\usepackage{url}
\usepackage{lineno}

\newcommand{\Aext}{\vec{\cal A}}
\newcommand{\Aextr}{{\cal A}_r}
\newcommand{\Aextps}{{\cal A}_1}
\newcommand{\Aextph}{{\cal A}_2}
\newcommand{\Bext}{\vec{\cal B}}
\newcommand{\Bextr}{{\cal B}_r}
\newcommand{\Bextps}{{\cal B}_1}
\newcommand{\Bextph}{{\cal B}_2}

\def\XXint#1#2#3{{\setbox0=\hbox{$#1{#2#3}{\int}$}
     \vcenter{\hbox{$#2#3$}}\kern-.5\wd0}}

\makeatletter
\g@addto@macro\bfseries{\boldmath}
\makeatother

\definecolor{nicered}{rgb}{0.7,0.1,0.1}
\definecolor{nicegreen}{rgb}{0.1,0.5,0.1}
\hypersetup{colorlinks, urlcolor=blue, citecolor=nicegreen,linkcolor= nicered}

\begin{document}
%\linenumbers
%TC: ignore
\title{Geomagnetic constraints on millicharged dark matter} 

\author{Ariel Arza}
\email{ariel.arza@gmail.com}
\affiliation{Department of Physics and Institute of Theoretical Physics, Nanjing Normal University, Nanjing, 210023, China}
\affiliation{Nanjing Key Laboratory of Particle Physics and Astrophysics, Nanjing, 210023, China}

\author{Yuanlin Gong}
\email{yuanlingong@nnu.edu.cn}
\affiliation{Department of Physics and Institute of Theoretical Physics, Nanjing Normal University, Nanjing, 210023, China}
\affiliation{Key Laboratory of Dark Matter and Space Astronomy, Purple Mountain Observatory,
Chinese Academy of Sciences, Nanjing 210008, China}

\author{Jing Shu}
\email{jshu@pku.edu.cn}
\affiliation{School of Physics and State Key Laboratory of Nuclear Physics and Technology, Peking University, Beijing 100871, China}
\affiliation{Center for High Energy Physics, Peking University, Beijing 100871, China}
\affiliation{Beijing Laser Acceleration Innovation Center, Huairou, Beijing, 101400, China}

\author{Lei Wu}
\email{leiwu@njnu.edu.cn}
\affiliation{Department of Physics and Institute of Theoretical Physics, Nanjing Normal University, Nanjing, 210023, China}
\affiliation{Nanjing Key Laboratory of Particle Physics and Astrophysics, Nanjing, 210023, China}

\author{Qiang Yuan}
\email{yuanq@pmo.ac.cn}
\affiliation{Key Laboratory of Dark Matter and Space Astronomy, Purple Mountain Observatory,
Chinese Academy of Sciences, Nanjing 210008, China}
\affiliation{School of Astronomy and Space Science, University of Science and Technology of China,
Hefei 230026, China}

\author{Bin Zhu}
\email{zhubin@mail.nankai.edu.cn}
\affiliation{School of Physics, Yantai University, Yantai 264005, China}

\begin{abstract}

Millicharged particles are well-motivated dark matter candidates arising in many extensions 
of the Standard Model. We show that, despite their tiny coupling $e_m$ to photons, 
millicharged dark matter (mDM) in the Earth's geomagnetic field can generate 
a quasi-static, monochromatic magnetic signal with angular frequency twice the mDM mass. 
Using null results from the SuperMAG and SNIPE Hunt collaborations, 
we constrain the effective charge of bosonic mDM in the mass range 
$10^{-18}$--$10^{-14}\,\text{eV}$. The resulting upper bounds exceed stellar cooling 
constraints by over thirteen orders of magnitude, demonstrating the power of this method. 

\end{abstract}

\maketitle
\newpage

\begin{figure*}[ht]
\centering
\includegraphics[width=0.8\linewidth]{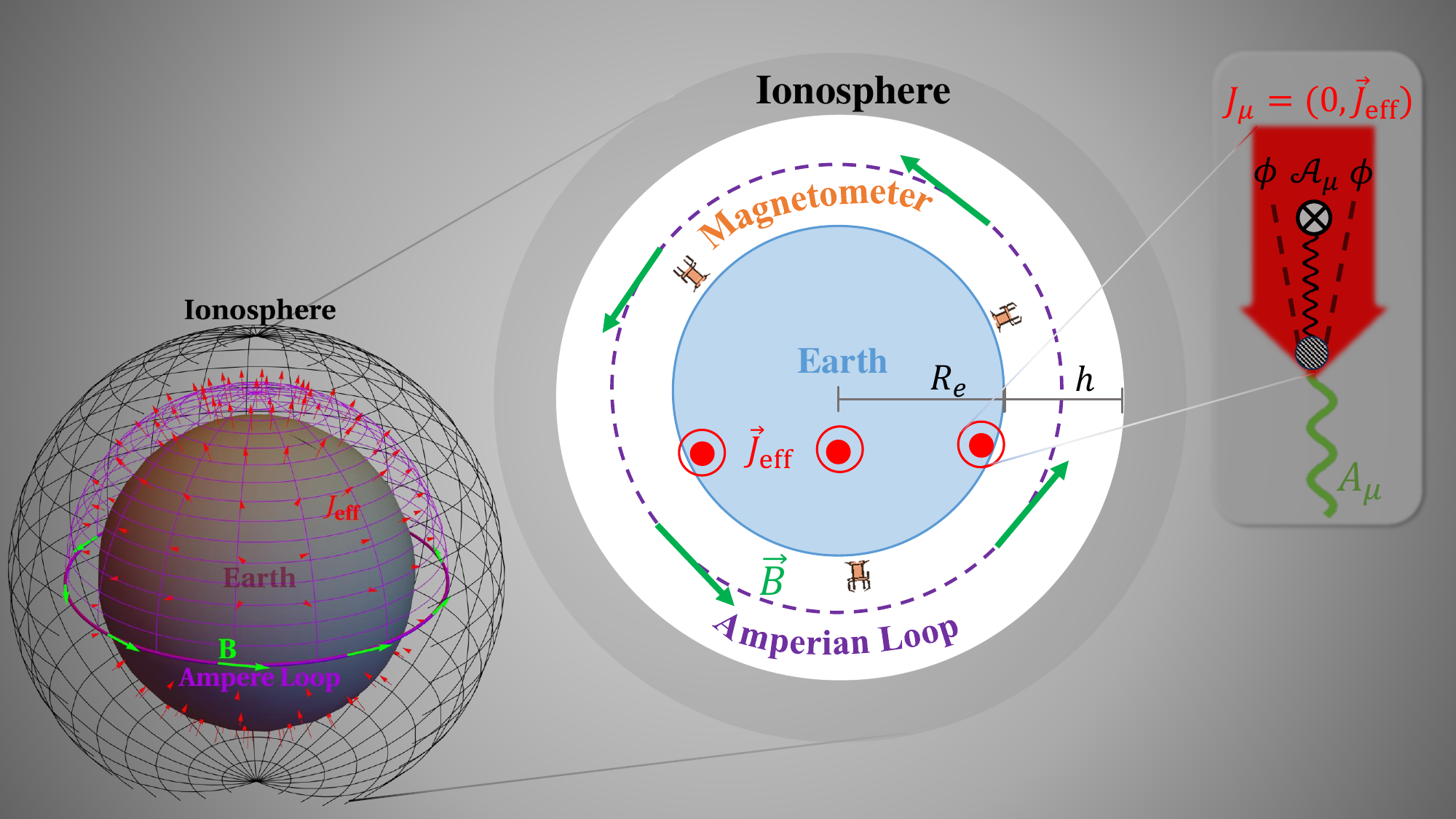}
\caption{\textbf{Middle}: Natural Earth's cavity formed between the  Earth's surface and the ionosphere. The radial component of the effective current $\vec{J}_{\mathrm{eff}}$ induced by mDM passing through a chosen Amperian Loop. The generated magnetic field $\vec{B}$ can be probed by the magnetometers (orange symbols) placed over the surface of the Earth. \textbf{Lower Left}: The 3D version of the Middle, where the effective current is depicted from Eq. (\ref{eq:Jeff1}). \textbf{Upper Right}: The signal magnetic field produced by the millicharged effective current in the geomagnetic field background.}
\label{setup}
\end{figure*}

{\it \textbf {Introduction}.} Identifying the nature of dark matter has been one of the frontiers in modern physics. Millicharged particles, i.e., quantum fields with electric charges much smaller than the one $e$ of the electron, are natural candidates \cite{Dimopoulos:1989hk,DeRujula:1989fe,Feldman:2007wj,McDermott:2010pa,Cline:2012is}. Millicharged dark matter (mDM) can arise from models where the visible and a dark sectors are kinetically mixed \cite{Holdom:1985ag, Goldberg:1986nk,Izaguirre:2015eya,Feldman:2007wj,Cheung:2007ut,Feng:2023ubl}, or from models where these particles are directly charged under the $U(1)_Y$ in the UV theory \cite{Wen:1985qj, Shiu:2013wxa, Feng:2014eja}. They are also widely predicted in string theory compactifications \cite{Wen:1985qj,Burgess:2008ri, Goodsell:2009xc, Cicoli:2011yh, Shiu:2013wxa, Feng:2014eja}, and grand unification theories \cite{Pati:1973uk, Georgi:1974my, Preskill:1984gd}. 

The masses of mDM can span in a wide range. Depending on their masses, mDM could have been produced in the early universe either via the freeze-in mechanism~\cite{Hall:2009bx,Dvorkin:2019zdi,Bhattiprolu:2023akk,Bhattiprolu:2024dmh,Essig:2011nj,Chu:2011be} or the misalignment mechanism~\cite{Preskill:1982cy,Abbott:1982af,Dine:1982ah,Nelson:2011sf,Arias:2012az}. In particular, the bosonic mDM can be of a low mass (possibly even sub-eV)~\cite{Alonso-Alvarez:2019pfe,Jaeckel:2021xyo}, which might offer a solution to the small-scale structure problems and be relevant to astrophysical observations~\cite{Hu:2000ke, Hui:2016ltb, Ferreira:2020fam}. For example, mDM with masses on the order of a few tens of MeV could account for the anomaly in the 21-cm signal if a small fraction of the dark matter is composed of them~\cite{Munoz:2018pzp,Berlin:2018sjs,Slatyer:2018aqg,Kovetz:2018zan,Liu:2019knx}.

There have been various experiments and proposals capable to search for mDM. They include cavities filled with strong electric field to probe Schwinger pair production \cite{Gies:2006hv, Berlin:2020pey,Romanenko:2023irv}, observation of the invisible decays of positronium \cite{Badertscher:2006fm}, lamb shift of the hydrogen atom \cite{Gluck:2007ia}, laser polarization experiments \cite{Gies:2006ca,Ahlers:2007qf,DellaValle:2014xoa, DellaValle:2015xxa}, Cavendish experiment for testing Coulomb's Law \cite{Jaeckel:2009dh}, timing of radio waves from pulsars and fast radio bursts \cite{Caputo:2019tms,liuyuxin}, millicharged condensates~\cite{Berlin:2024dwg}, and direct deflection type experiments \cite{Berlin:2019uco, Berlin:2021kcm, Berlin:2023gvx}. Meanwhile, astrophysical observations including the SN1987A \cite{Mohapatra:1990vq,Chang:2018rso,Fiorillo:2024upk} and stellar evolution \cite{Bernstein:1963qh,Dobroliubov:1989mr,Davidson:1991si,Davidson:2000hf,Vogel:2013raa,Fung:2023euv} have placed very strong constraints, and even more stringent ones can be obtained from galactic large scale magnetic fields \cite{Stebbins:2019xjr} and cosmological measurements \cite{Davidson:1993sj,Melchiorri:2007sq,Burrage:2009yz,Jaeckel:2021xyo}, although the latter are highly model-dependent (for comprehensive reviews, see also \cite{Bogorad:2021uew,Jaeckel:2010ni}).

In this work, we propose a new method to search for ultralight mDM by detecting their magnetic signal on Earth’s surface. For sub-eV masses, the high occupation number allows the mDM field to be treated as a classical wave. When the Compton wavelength is much larger than the Earth's radius, the mDM field coherently interacts with the geomagnetic field, producing a quasi-static, monochromatic magnetic signal with angular frequency equal to twice the mDM mass, and width given by $\left<v_\phi^2\right>\sim10^{-6}$ of the main signal frequency, where $v_\phi$ is the mDM velocity. It is important to emphasize that the magnetic field induced by mDM shows a characteristic inverse-square dependence on the mass in the relevant parameter regime,
\begin{equation}
B_{\phi} \sim {4 e_m^2 B_0 R_e^2\, \rho}/{m_\phi^2} ~~ ,
\end{equation}
where $e_m$ is the millicharge, $m_\phi$ the mDM mass, $B_0$ the geomagnetic field, $R_e$ the Earth's radius, and $\rho$ the local dark matter energy density. This strong $1/m_{\phi}^2$ scaling significantly enhances the sensitivity to ultra-light mDM. By contrast, other dark-matter candidates induce magnetic signals with different mass dependence:
\begin{itemize}
    \item Dark photons: $B_{A'} \sim \epsilon m_{A'} R_e \sqrt{2\rho}$~\cite{Fedderke:2021aqo, Fedderke:2021rrm}
    \item Axions: $B_a \sim g_{a\gamma} B_0 R_e \sqrt{2\rho}$~\cite{Arza:2021ekq}
\end{itemize}
This renders mDM very distinct and promising in searches for ultra-light dark matter.

By using null results for axion and dark photon searches from the SuperMAG \cite{Fedderke:2021aqo,Fedderke:2021rrm,Arza:2021ekq,Friel:2024shg} and SNIPE Hunt \cite{Sulai:2023zqw} collaborations, we estimate a world-leading bound on ultralight mDM, which surpasses the existing constraints from stellar cooling by more than thirteen orders of magnitude. Consequently, it is expected that future dedicated measurements of the magnetic activity on the Earth's surface will open up new avenues for the exploration of mDM on different mass ranges and smaller values of $e_m$.

{\it\textbf{mDM Electrodynamics}.} The interaction between the ultralight bosonic mDM field $\phi$ and the photon field $A_\mu$ is described by the following lagrangian
\begin{align}
{\cal L}=D_\mu\phi(D^\mu\phi)^*-m_\phi^2|\phi|^2-{\frac{1}{4}}F_{\mu\nu}F^{\mu\nu} ~~ , \label{eq:lag1}
\end{align}
where $D_\mu=\partial_\mu+ie_mA_\mu$ is the covariant derivative and $F_{\mu\nu}=\partial_\mu A_\nu-\partial_\nu A_\mu$ is the electromagnetic strength tensor\footnote{Recent studies~\cite{Ioannisian:2017srr,Beutter:2018xfx} have calculated the transition probability between ultralight dark matter and photons using quantum field theory, consistent with earlier results from classical Maxwell equations. For convenience, we derive the induced electromagnetic signal from the modified classical Maxwell equations.}. For a given external electromagnetic field ${\cal A}_\mu=({\cal A}_0,-\vec{\cal A})$, we make the decomposition $A_\mu\rightarrow A_\mu+{\cal A}_\mu$, where now $A_\mu$ corresponds explicitly to the induced electromagnetic signal. The equations of motion are then written as
\begin{eqnarray}
\partial_\mu F^{\mu\nu} +2e_m^2|\phi|^2A^\nu &=& J_m^\nu-2e_m^2{\cal A}^\nu|\phi|^2 \equiv J^\nu_\text{eff}, \label{eq:max1}  \\ 
(\Box+m_\phi^2) \phi&=&-ie_m\partial_\mu {\cal A}^\mu\phi-2ie_m\partial_\mu\phi {\cal A}^\mu \nonumber \\ && +e_m^2{\cal A}_\mu {\cal A}^\mu\phi ~~ , \label{eq:phi1}
\end{eqnarray}
where $J_m^\nu=ie_m(\phi^*\partial^\nu\phi-\phi\,\partial^\nu\phi^*)$ is the millicharged current and the non-linear terms were neglected in Eq. \eqref{eq:phi1}. In Eq. (\ref{eq:max1}), the effective mass term $2e_m^2|\phi|^2A^\nu$ can, in principle, induce a photon parametric resonance \cite{Jaeckel:2021xyo}. However we find that this effect is subdominant in the parameter space of our interest, and can safely neglect it.

Gauge invariance persists in our model at all orders. Indeed, it is evident from the equations of motion, as the above mass term in Eq.~(\ref{eq:max1}) and the non-linear terms in Eq.~(\ref{eq:phi1}) were neglected. Even when the mass term is taken into account, the gauge invariance is realized from the significant back-reaction on the dark matter field once the previously neglected nonlinear terms in Eq. (\ref{eq:phi1}) are taken into account. For convenience, we adopt the Coulomb gauge, i.e., $\vec\nabla\cdot\vec A=0$ and $A_0=0$.

Once we know the mDM field $\phi$ at the Earth's location, the electromagnetic signal is found by solving Eq. (\ref{eq:max1}).

Due to the high occupation number of sub-eV mDM, it is typically treated as a free classical wave, 
$\phi = \phi_0 = {\sqrt{2\rho}\over m_\phi} \cos(\vec{k}_\phi \cdot \vec{x} - m_\phi t)$, 
where $\vec{k}_\phi=m_\phi\vec v_\phi$ denotes the mDM momentum. However, as the mDM field approaches the Earth, 
the Earth's geomagnetic field can alter the field configuration, especially for 
ultralight masses. To capture this effect, we solve Eq.~(\ref{eq:phi1}) with the boundary 
condition $\phi \rightarrow \phi_0$ as $|\vec{x}| \rightarrow \infty$. For the mass range 
relevant to our study, the de Broglie wavelength of mDM greatly exceeds the Earth's radius. 
This allows us to simplify the problem by modeling the Earth's effect as a Dirac delta 
potential localized at $\vec{x} = 0$, leading to the solution~\cite{Jackiw:1991je}
\begin{equation}
\phi(t)= {1\over2}{\sqrt{2\rho}\over m_\phi}(1+i\varepsilon)\left({e^{-i\omega_\phi t}\over1+i\kappa\varepsilon}+{e^{i\omega_\phi t}\over1-i\kappa\varepsilon}\right) ~~, \label{eq:phigen}  
\end{equation}
where $\omega_\phi = \sqrt{k_\phi^2 + m_\phi^2} \simeq m_\phi$ is the mDM energy, and we define 
two dimensionless parameters, $\kappa \approx {e_m B_0 R_e}/{k_\phi}$ and 
$\varepsilon \approx 8 e_m B_0 R_e^2$. A small $\kappa$ results in a large gyroradius so that the trajectories of the particles are barely deflected by Earth's magnetic field. Meanwhile, a small $\varepsilon$ highlights the wave-like behavior of the millicharged particles, leading to significant tunneling through the geomagnetic potential barrier that would otherwise repel them. When the interaction with the Earth's magnetic field is weak, 
i.e., $\varepsilon, \kappa \ll 1$, Eq.~(\ref{eq:phigen}) reduces to the free-wave form 
$\phi = \phi_0 \sim \sqrt{2\rho}/m_\phi$. 
For the parameter region where we set upper bounds on $e_m$, these conditions are always satisfied. 
In this regime, the effective current $J_{\text{eff}} \sim e_m^2 \mathcal{A} |\phi|^2$ 
scales as $1/m_\phi^2$, enhancing the signal at small $m_\phi$.

On the other hand, when $m_\phi$ is very small such that $\kappa \gg 1$, the $1/m_\phi$ scaling in 
$\phi$ is canceled, making $J_{\text{eff}}$ finite even as $m_\phi \to 0$. 
If the mDM electric charge is large, i.e., $\varepsilon, \kappa \gg 1$, 
Eq.~(\ref{eq:phigen}) simplifies to 
$\phi = -i \sqrt{2\rho} \, v_\phi / (e_m B_0 R_e) \sin(m_\phi t)$. In this limit, the mDM is repelled by the geomagnetic potential, 
and the effective current saturates at 
$J_{\text{eff}} \sim \rho v_\phi^2 / (B_0 R_e)$, 
independent of $m_\phi$ and $e_m$, corresponding to a constant magnetic signal of 
$1.43\,\mathrm{pT}$ in our study. Further discussion of the mDM wave function behavior is provided in the Supplementary Material.

{\it \textbf{Experimental setup and sensitivity}.} Let us consider a region of size $L$ that is bounded by a conducting shielding. A dark matter induced oscillating effective current $J_\text{eff}$ drives an electromagnetic field inside the region that is damped in the boundaries. When the frequency of the oscillating current is much smaller than $1/L$, the signal is dominated by a quasi-static magnetic field whose strength is roughly of the form $B\sim J_\text{eff}L$. Based on this, novel experimental ideas for axions \cite{Sikivie:2013laa,DMRadio:2022pkf} and also dark photons \cite{Arias:2014ela} were proposed at the laboratory scale for masses between $10^{-10}$ and $10^{-6}\,\text{eV}$, where one can benefit from resonant signal enhancements and strict control of the background noise.

In this work, we propose to search for bosonic mDM by applying the same principle mentioned above to the huge natural Earth's conducting cavity. As illustrated in Fig.~\ref{setup}, the Earth can play a role of a natural cavity, shielded from below by the Earth's surface and from above by the ionosphere (or interplanetary medium). We set the Earth's radius as $R_e=6371.2\,\text{km}$, therefore the proposal applies for masses  below $1/R_e\sim10^{-14}\,\text{eV}$. For such low masses, we benefit in sensitivity due to the Earth size and the $1/m_\phi^2$ scaling of the signal. Using high precision magnetometers situated over the surface of the Earth, it is possible to probe ultralight mDM in a wide unexplored parameter space.

The profile of the vector potential $\Aext$ above the Earth's surface is determined by the profile of the geomagnetic field $\Bext$ extended over the whole Earth's region. We use the updated IGRF model \cite{alken2021international} to describe the geomagnetic field from the atmosphere into the mantle, while for the inner and outer core, we model it by resorting to the electric current that produces it. The most popular theory about the origin of the Earth's magnetic field is the geodynamo, in which the geomagnetic field is generated by electric currents $\vec{J}$ in the Earth's outer core~\cite{1995PEPI...91...63G,1995Natur.377..203G}. Currents in the solid inner core and the mantle are vanishing. To get the profile of $\vec{J}$, we use polynomial functions that fit the rms value of the geomagnetic field $\cal B$ over the outer core and respect appropriate boundary conditions. Then, the magnetic field $\Bext$ and vector potential $\Aext$ can be calculated by solving $\nabla\times\Bext = \vec{J}$ and $\nabla\times\Aext = \Bext$, respectively, together with the conditions $\vec\nabla\cdot\Bext=0$ and $\vec\nabla\cdot\Aext=0$. See all the details of this calculation in Supplementary Material.

As a result, the external vector potential above the Earth's surface, i.e., $r\geq R_e$, is given by
\begin{align}
\Aext = & -B_0R_e\left(1.01\beta\left(R_e\over r\right)^3(2\vec Y_{10}-\vec\Psi_{10})\right. \nonumber
\\
& \ \ \ \ \ \ \ \left.-\sqrt{4\pi\over3}\left(R_e\over r\right)^2\vec\Phi_{10}\right) ~~ , \label{eq:Aextsol}
\end{align}
where $B_0=2.94\times10^{-2}\,\text{mT}$ is the total geomagnetic field at the equator of the Earth's surface and $\vec Y_{10},\; \vec\Psi_{10},\; \vec\Phi_{10}$ are vector spherical harmonics (VSH). The parameter $\beta$ accounts for the uncertainty in the rms value of the geomagnetic field ${\cal B}$ in the outer core. The value of $\beta$ ranges between 1 and 4.6.

\begin{figure}[t]
\centering
\includegraphics[width=0.98\linewidth,height=7cm]{ 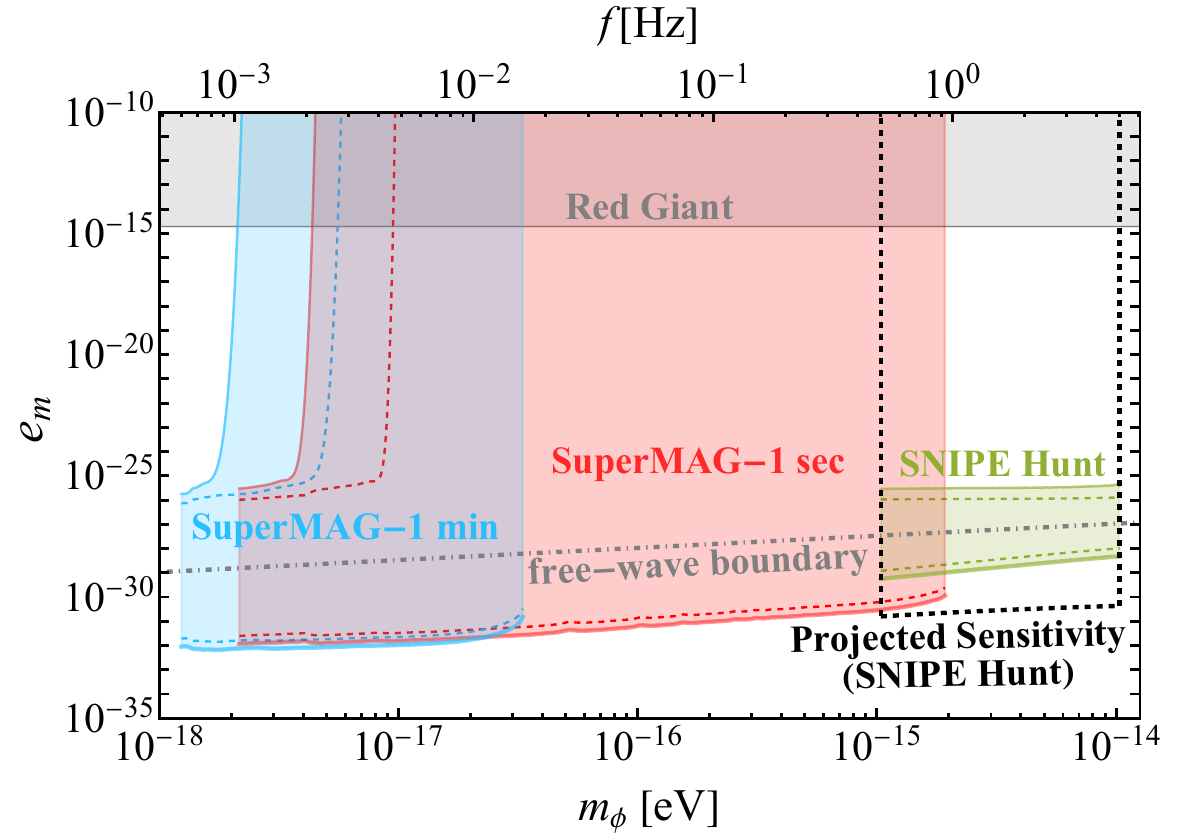}
\caption{Sensitivity for millicharged dark matter in the plane of $m_\phi$ and $e_m$. The blue, red and green shaded regions are excluded by the null results from the SuperMAG-1min, SuperMAG-1sec and SNIPE Hunt experiments, respectively. The dashed (solid) lines correspond to $\beta=1$ (4.6) reflecting the uncertainty on the geomagnetic field rms value over the Earth's outer core. The lower edge of our constraint 
is well below the free-wave boundary 
$\kappa = 1$ (grey dash-dotted line), as given by Eq.~(\ref{eq:Bsignal2_fin}). 
We also show the bound from red giant cooling~\cite{Fung:2023euv} and the projected sensitivity 
of the SNIPE Hunt experiment using a LEMI-120 sensor with one week of data~\cite{SNIPEhunt2025}. 
This projected reach fully covers the previously 
unconstrained region, leaving no lower limit in the parameter space.}
\label{fig:sens}
\end{figure}

Knowing the external vector potential $\Aext$, we can analytically solve Eq.~(\ref{eq:max1}). For our mass range of interest, where $m_\phi \ll 1/R_e$ is satisfied, Eq. (\ref{eq:max1}) can be approximated to the magneto quasi-static equation
\begin{equation}
\vec\nabla\times\vec B=\vec J_\text{eff} ~~ , \label{eq:Bsignal1}
\end{equation}
where $\vec B=\vec\nabla\times\vec A$ is the magnetic field signal and $\vec J_\text{eff}$ the dark matter effective current given by
\begin{align}
\vec J_\text{eff} = & {2e_m^2\,\rho B_0R_e\over m_\phi^2}\left(1.01\beta\left(R_e\over r\right)^3(2\vec Y_{10}-\vec\Psi_{10})\right. \nonumber
\\
& \ \ \ \ \ \ \ \left.-\sqrt{4\pi\over3}\left(R_e\over r\right)^2\vec\Phi_{10}\right)e^{-2i m_\phi t} ~~ . \label{eq:Jeff1}
\end{align}

In our calculation, we use the fact that both the Earth's surface and the ionosphere layers are well conductors. Then, Eq.~(\ref{eq:Bsignal1}) can be solved with the boundary condition $\vec B_r=0$ at $r=R_e$ and $r=R_e+h$, where $\vec B_r$ is the radial component of the magnetic field and $h$ is the ionosphere height. We note that, at very low frequencies and depending on solar activity and other factors, the ionosphere may not be a good conductor that can damp the electromagnetic active mode completely. In such a case the upper boundary would be the interplanetary medium. For more details of the ionosphere and interplanetary medium conducting properties, see Ref. \cite{Fedderke:2021aqo}.

Finally, by solving Eq. (\ref{eq:Bsignal1}) together with the condition $\vec\nabla\cdot\vec B=0$, the magnetic field signal at the Earth's surface is given by
\begin{align}
\vec B(t) = & {e_m^2\,\rho B_0R_e^2\over m_\phi^2}\left(\sqrt{4\pi\over3}{h\over R_e}\vec\Psi_{10}\right. \nonumber
\\
& \ \ \ \ \ \left.-2.02\beta\,\vec\Phi_{10} \right)e^{-i2m_\phi t} ~~ . \label{eq:Bsignal2}
\end{align}
The dominant contribution comes from the $\vec\Phi_{10}$ term of the signal. The suppression factor $h/R_e \ll 1$ in the $\vec\Psi_{10}$ term can be understood from the Ampere's law $\oint d\vec l\cdot\vec B=\int d\vec S\cdot\vec J_\text{eff}$, where an integral loop is shown in Fig.~S2 in the Supplementary Material. For pure tangential components of the effective current in Eq. (\ref{eq:Jeff1}), i.e. the $\vec\Phi_{10}$ term, the integral $\int d\vec S\cdot\vec J_\text{eff}$ gives $\sim \pi R_ehJ_\text{eff}$, while $\oint d\vec l\cdot\vec B$ becomes $\sim BR_e$, resulting in $|B| \sim \pi R_e(h/R_e)J_\text{eff}$. On the other hand, for radial components of the effective current, such as the $\vec Y_{10}$ term in Eq. (\ref{eq:Jeff1}), we can use the amperian loop shown in Fig. \ref{setup} to obtain $\int d\vec S\cdot\vec J_\text{eff}\sim\pi R_e^2J_\text{eff}$, leading to $B\sim\pi R_eJ_\text{eff}$.

For $R_e = 6371.2~\text{km}$,  $B_0 = 2.94\times 10^{-2}\,\text{mT}$, $\beta = 1$ and $\rho= 0.3\,\rm{GeV/cm^3}$, we notice that the amplitude of the magnetic field signal is 
\begin{equation}
|B|=696\,\text{pT}\left(e_m\over10^{-27}\right)^2\left(10^{-14}\,\text{eV}\over m_\phi\right)^2 ~~ , \label{eq:Bsignal2_fin}
\end{equation}
meaning that for our mass ranges of interest, which correspond to signal frequencies below $5~\text{Hz}$, a mDM field can produce a magnetic field signal of hundreds of pT for millicharges down to $e_m = 10^{-27}$. Current unshielded magnetometers can easily reach these sensitivities. To have an idea, in the first campaign of the SNIPE Hunt collaboration, magnetometers with $300\,\text{pT}/\sqrt{\text{Hz}}$ were used \cite{Sulai:2023zqw}.

A straightforward concern for a dedicated experimental exploration are the background noises. For the frequencies of interest, the dominant magnetic field interference is of anthropogenic origin \cite{Constable}. Therefore, to get optimal sensitivity, the experiment can be set up in a location far away from urban environments. Moreover, the mDM signal oscillates at a particular frequency that makes it easy to be distinguished from static sources, such as the strong static Earth's geomagnetic field. In fact, it was demonstrated in the first SNIPE Hunt expedition, that the amplitude spectral density of the measurements were flat and corresponded to the noise floor of the magnetometers, with the exception of a 60 Hz peak associated with the laptop used for data acquisition. While a few narrow peaks at one of the measurement stations are attributed to noise with unknown origin, our signal should be present in all stations at all times. In addition, since the reach of the SNIPE Hunt is now limited by the sensitivity of the magnetometers \cite{Sulai:2023zqw}, rather than by the geomagnetic noise, future dedicated experiments using high-precision magnetometers with sensitivity up to order $20\,\text{fT}/\sqrt{\text{Hz}}$ \cite{Bevilacqua:2016zpi,Oelsner:2020lcl,Chatzidrosos:2017tpj,Younesi:2024azs} will result in stronger sensitivity. Moreover, methods to improve sensitivity and frequency range coverage have already being examined. For instance, it has been pointed out in \cite{Bloch:2023wfz} that the measurement of the curl of the magnetic field can serve as a useful method to reach much higher frequencies.

In Fig.~\ref{fig:sens}, we show constraints for mDM by recasting the null results of searching for axion and dark photon from the SuperMAG \cite{Fedderke:2021aqo,Fedderke:2021rrm,Arza:2021ekq,Friel:2024shg} and SNIPE Hunt \cite{Sulai:2023zqw} collaborations. The SNIPE Hunt collaboration performed coordinated measurements with a network of magnetometers located in the US, in three different stations far from human-generated magnetic noise. Each station used a vector magnetoresistive sensor with sensitivity of 300 $\mathrm{pT}/\sqrt{\mathrm{Hz}}$ over a frequency range of $0.1-100$ Hz \cite{Sulai:2023zqw}. The SuperMAG data consist of two sets of measurements taken over several tens of years. One set was conceived at 1-minute intervals with contributions from $\mathcal{O}(500)$ stations all over the world \cite{Arza:2021ekq, Fedderke:2021aqo, Fedderke:2021rrm}. The other set corresponds to measurements with a 1-second time resolution with contributions from $\mathcal{O}(200)$ stations. This ``high-fidelity'' dataset was analyzed for the cases of dark photon and axion dark matter in Ref. \cite{Friel:2024shg}. To obtain the SNIPE Hunt constraints for mDM, we first calculate the magnetic field amplitude $|\vec{B}_a|$ by using the constraints found in Ref.~\cite{Sulai:2023zqw} for three different stations and then obtain an averaged value. In the same frequency, we equal this average to the counterpart magnetic field signal in Eq.~(\ref{eq:Bsignal2}) to estimate the bounds on $e_m$. For the SuperMAG constraints, since the data are collected with many stations situated over the world, we instead average the $\vec{B}_a$ over the Earth's surface to get magnetic field expressions independent on the Earth location. Then we proceed in the same way as done for SNIPE Hunt. As we can see in Fig.~\ref{fig:sens}, the estimated constraints surpass previous bounds from stellar cooling by over thirteen orders of magnitude in the case of SNIPE Hunt and over seventeen orders of magnitude for SuperMAG.

We note that regions with large $e_m$, where mDM is repelled significantly by the geomagnetic potential, 
remain unconstrained by current SuperMAG and SNIPE Hunt data. 
This is because, at the edge of our sensitivity window for $m_\phi$, the minimum detectable signal 
exceeds the saturation value of $1.43\,\mathrm{pT}$, as discussed below Eq.~(\ref{eq:phigen}).
As a result, mDM with sufficiently large charges becomes undetectable in this regime. These lower bounds are derived using the wave function from 
Eq.~(\ref{eq:phigen}) in the signal estimation. Still, within the relevant mass range, enhancing the magnetometer sensitivity by a factor of three 
would suffice to fully cover the large-charge parameter space. This points to a realistic and 
promising direction for future experimental efforts to close the remaining gap in the mDM search. 
As shown in Fig.~\ref{fig:sens}, the projected sensitivity of a high-performance LEMI-120 
induction coil magnetometer, operating in the $0.5 \sim 5\,\mathrm{Hz}$ range for one week, 
can reach this level~\cite{SNIPEhunt2025}.

{\it \textbf{Conclusion and outlook}.} We have presented a novel approach to probe ultralight mDM by measuring their geomagnetic signals. Using classical field theory, we calculated the quasi-static monochromatic magnetic field sourced by the effective current associated with the mDM field, which exhibits a distinctive inverse-square dependence on the mDM mass. By reinterpreting the data from the SuperMAG and SNIPE Hunt experiments, we have derived stringent constraints on the effective charge of bosonic mDM over specific mass ranges. These bounds exceed stellar cooling limits by over thirteen and up to eighteen orders of magnitude. 
For two narrow mass windows, unconstrained regions remain at large $e_m$ due to the repulsion of mDM by the geomagnetic potential. Nonetheless, a factor of three improvement in magnetometer sensitivity would be sufficient to close these gaps.
Finally, a global network of high-precision magnetometers monitoring Earth's 
magnetic field could offer a powerful platform for ultralight mDM detection 
in the future.

{\it \textbf{Acknowledgement}}. We would like to thank Haipeng An, Itay Bloch, Wei Ji, Saarik Kalia and Yuxin Liu for useful discussions. L.W. is supported by the National Natural Science Foundation of China (NSFC) under Grants No. 12275134 and No. 12335005. B.Z. is supported by the NSFC under Grant No. 12275232. J.S. is supported by the National Key Research and Development Program of China under Grants No. 2020YFC2201501 and No. 2021YFC2203004, Peking University under startup Grant No. 7101302974, the NSFC under Grants No. 12025507, No. 12150015, No. 12450006, and the Key Research Program of Frontier Science of the Chinese Academy of Sciences (CAS) under Grant No. ZDBS-LY-7003. Q.Y. is supported by the NSFC under Grant No. 12220101003 and the Project for Young Scientists in Basic Research of the CAS under Grant No. YSBR-061.

{\it \textbf{Data availability}}. The datasets generated during and/or analysed during the current study are available from the corresponding author on request.

{\it \textbf{Code availability}}.
The custom computer codes used to generate results are available
from the corresponding author on request.

\bibliography{refs_prl}

%\documentclass[aps,prl,longbibliography,superscriptaddress,amsfont,graphicx,nofootinbib,preprintnumbers]{revtex4-1}%
%\UseRawInputEncoding
%\usepackage{color,graphicx,epsfig}
%\usepackage{ifpdf}
%\usepackage{amsmath}
%\usepackage{bm}
%\usepackage[english]{babel}
%\usepackage{amssymb}
%\usepackage{braket}
%\usepackage{setspace}
%\allowdisplaybreaks[4]

%\usepackage[colorlinks,linkcolor=blue,anchorcolor=blue,citecolor=blue,urlcolor=blue]{hyperref}
%\usepackage{hyperref}
%\usepackage{enumerate}
%\usepackage{url}
%\usepackage{multicol}
%\usepackage{lipsum}
%\usepackage{comment}

%\bibliographystyle{apsrev}

%\usepackage{slashed}

%\newcommand{\SLASH}[2]{\makebox[#2ex][l]{$#1$}/}
%\usepackage{changes}
%\definechangesauthor[name={Liangliang Su}, color=orange]{LLS}
%\setremarkmarkup{(#2)}

%\begin{document}

\onecolumngrid
\clearpage

%%%%%%%%%% Supplemental materials %%%%%%%%%%
\setcounter{page}{1}
\setcounter{equation}{0}
\setcounter{figure}{0}
\setcounter{table}{0}
\setcounter{section}{0}
\setcounter{subsection}{0}
\renewcommand{\theequation}{S.\arabic{equation}}
\renewcommand{\thefigure}{S\arabic{figure}}
\renewcommand{\thetable}{S\arabic{table}}
\renewcommand{\thesection}{\Roman{section}}
\renewcommand{\thesubsection}{\Alph{subsection}}

\newcommand{\ssection}[1]{
    \addtocounter{section}{1}
    \section{\thesection.~~~#1}
    \addtocounter{section}{-1}
    \refstepcounter{section}
}
\newcommand{\ssubsection}[1]{
    \addtocounter{subsection}{1}
    \subsection{\thesubsection.~~~#1}
    \addtocounter{subsection}{-1}
    \refstepcounter{subsection}
}
\newcommand{\fakeaffil}[2]{$^{#1}$\textit{#2}\\}

\thispagestyle{empty}
\begin{center}
    \begin{spacing}{1.2}
        \textbf{\large
            \hypertarget{sm}{Supplemental Material:} Geomagnetic constraints on millicharged dark matter}\\
    \end{spacing}
    \par\smallskip
    Ariel Arza,$^{1,2}$
    Yuanlin Gong,$^{1,3}$
    Jing Shu,$^{4,5,6}$
    Lei Wu,$^{1,2}$
    Qiang Yuan,$^{3,7}$
    Bin Zhu$^{8}$
    \par
    {\small
        \fakeaffil{1}{Department of Physics and Institute of Theoretical Physics, Nanjing Normal University, Nanjing, 210023, China}
        \fakeaffil{2}{Nanjing Key Laboratory of Particle Physics and Astrophysics, Nanjing, 210023, China}
        \fakeaffil{3}{Key Laboratory of Dark Matter and Space Astronomy, Purple Mountain Observatory, Chinese Academy of Sciences, Nanjing 210023, China}
        \fakeaffil{4}{School of Physics and State Key Laboratory of Nuclear Physics and Technology, Peking University, Beijing 100871, China}
        \fakeaffil{5}{Center for High Energy Physics, Peking University, Beijing 100871, China}
        \fakeaffil{6}{Beijing Laser Acceleration Innovation Center, Huairou, Beijing, 101400, China}
        \fakeaffil{7}{School of Astronomy and Space Science, University of Science and Technology of China, Hefei 230026, China}
        \fakeaffil{8}{School of Physics, Yantai University, Yantai 264005, China}
    }
\end{center}
\par\smallskip

This Supplemental Material provides detailed calculations for the conversion of mDM into 
magnetic signals in the Earth's geomagnetic field. We begin by deriving the mDM wave function, 
followed by the formal solution for the associated vector potential. Finally, we present a 
model for the geomagnetic field and compute the resulting vector potential profile.

\section{Wave Function of mDM  }

In this section we consider the impact of the Earth's geomagnetic field on the incoming mDM wave. We consider a well localized Dirac delta potential, which is well justified by the fact that the mDM wavelength is way longer than the potential extent.

The effects of the Earth's geomagnetic field on the mDM can be found by solving Eq. (\ref{eq:phi1}). We write the mDM field in the general form
\begin{equation}
\phi(\vec x,t)=\phi_+(\vec x)e^{-i\omega_\phi t}+\phi_-(\vec x)e^{i\omega_\phi t} ~~ ,
\end{equation}
where $\omega_\phi=\sqrt{k_\phi^2+m_\phi^2}$. After replacing this ansatz and imposing the gauge conditions into Eq. (\ref{eq:phi1}), it becomes
\begin{equation}
(\nabla^2+k_\phi^2)\phi_{\pm}=2ie_m\vec{\cal A}\cdot\vec\nabla\phi_{\pm}+e_m^2{\cal A}^2\phi_{\pm} ~~ . \label{eq:phipmeqd1}
\end{equation}

Since the de Broglie wavelength of the ultralight mDM is much larger than the Earth's size and the spatial extent of the effective potential, we can approximate the potential as a Dirac delta function %We assume a vector potential of the form
\begin{equation}
\vec{\cal A}(\vec x)=\vec {\cal A}_0V\delta^3(\vec x) ~~ , \label{eq:Ad1}
\end{equation}
where ${\cal A}_0\sim B_0R_e$ and $V$ a typical interaction volume of the potential. The value of this volume can be inferred as $V\sim \sigma L$, where $\sigma$ can be identified as the total interaction cross section of the potential and $L$ the spatial extent of the potential in the direction of the wave vector $\vec k_\phi$. We now write the field $\phi_\pm$ as
\begin{equation}
\phi_\pm(\vec x)={1\over2}{\sqrt{2\rho}\over m_\phi}e^{\pm i\vec k_\phi\cdot\vec x}+\phi_s^\pm(\vec x) ~~ , \label{eq:phipmd1}    
\end{equation}
where the first term corresponds to the incoming plane wave and the second the scattered wave. Replacing Eq. (\ref{eq:Ad1}) and Eq. (\ref{eq:phipmd1}) into Eq. (\ref{eq:phipmeqd1}), we get that $\phi_s^\pm$ can be obtained by solving
\begin{equation}
\left(\nabla^2+k_\phi^2\right)\phi_s^\pm=\left(2ie_m\vec{\cal A}_0\cdot\vec\nabla\phi_\pm(0)+e_m^2{\cal A}_0^2\,\phi_\pm(0)\right)V\delta^3(\vec x) ~~ ,
\end{equation}
where we use ${\cal A}(\vec x)^2= {\cal A}_0^2 V^2 \delta^3(0) \delta^3(\vec x) = {\cal A}_0^2 V\delta^3(\vec x)$.

We can find $\phi_s^\pm$ using the Green's function method, where for our case the Green's function is given by
\begin{align}
G_\pm(\vec x-\vec x\,') &=-\int{d^3q\over(2\pi)^3}{e^{i\vec q\cdot(\vec x-\vec x\,')}\over q^2-k^2\mp i\epsilon} \label{eq:Gd1}
\\
&=-{1\over4\pi}{e^{\pm ik_\phi|\vec x-\vec x'|}\over|\vec x-\vec x\,'|} ~~ . \label{eq:Gd2}
\end{align}
The full solution for $\phi_\pm$ is
\begin{align}
\phi_\pm(\vec x) &= {1\over2}{\sqrt{2\rho}\over m_\phi}e^{\pm i\vec k_\phi\cdot\vec x}+\int d^3x'G_\pm(\vec x-\vec x\,')\left(2ie_m\vec{\cal A}_0\cdot\vec\nabla\phi_\pm(0)+e_m^2{\cal A}_0^2\,\phi_\pm(0)\right)V\delta^3(\vec x\,') \nonumber
\\
&={1\over2}{\sqrt{2\rho}\over m_\phi}e^{\pm i\vec k_\phi\cdot\vec x}+\left(2ie_m\vec{\cal A}_0\cdot\vec\nabla\phi_\pm(0)+e_m^2{\cal A}_0^2\,\phi_\pm(0)\right)VG_\pm(\vec x) ~~ . \label{eq:phipmd2}
\end{align}
On the other hand, we notice from Eq. (\ref{eq:Gd1}) that $\vec\nabla G_\pm=0$, then
\begin{equation}
\vec\nabla\phi_\pm(\vec x)=\pm{i\vec k\over2}{\sqrt{2\rho}\over m_\phi}e^{\pm i\vec k_\phi\cdot\vec x} ~~ . \label{eq:gradphipmd2}
\end{equation}
We evaluate Eq. (\ref{eq:gradphipmd2}) at $\vec x=0$ and then replace into Eq. (\ref{eq:phipmd2}). Then we evaluate Eq. (\ref{eq:phipmd2}) at $\vec x=0$ to obtain an algebraic equation for $\phi_\pm(0)$. The solution for $\phi_\pm(0)$ is
\begin{equation}
\phi_\pm(0)={1\over2}{\sqrt{2\rho}\over m_\phi}\left({1\mp2e_m\vec{\cal A}_0\cdot\vec k\,VG_\pm(0) \over 1-e_m^2{\cal A}_0^2\,VG_\pm(0)}\right) ~~ . \label{eq:phipm0d1}
\end{equation}
We can notice in Eq. (\ref{eq:Gd1}) and Eq. (\ref{eq:Gd2}) that $G_\pm$ has a divergence at $\vec x=0$. This issue appears typically when solving a wave equation in a two or three dimensional Dirac delta potential. It can be addressed by renormalization of the coupling after integrating $q$ in Eq. (\ref{eq:Gd1}) up to certain cut off $\Lambda$ as done by Jackiw in \cite{Jackiw:1991je}. After absorbing the cut off in the coupling (coupling renormalization), the value for $G_\pm(0)$ is
\begin{equation}
G_\pm(0)=\mp {ik_\phi\over4\pi} ~~ .  
\end{equation}
We now can write $\phi_\pm(0)$ as
\begin{equation}
\phi_\pm(0) = {1\over2}{\sqrt{2\rho}\over m_\phi}{1+i\varepsilon \over 1\pm i\kappa\varepsilon} ~~ , \label{eq:phipm0d2}
\end{equation}
where $\varepsilon$ and $\kappa$ are real parameters defined as

\begin{align}
\varepsilon &= {1\over2\pi} e_m{\cal A}_0k_\phi^2\sigma L   \label{eq:epsd1}
\\
\kappa & = {e_m{\cal A}_0\over2k_\phi} ~~ , \label{eq:deltad1}
\end{align}
where for simplicity we have assumed a wave vector parallel to the potential direction. We can easily identify $L$ as $2R_e$ while the interaction cross section $\sigma$ needs further analysis. To get $\sigma$ we write Eq. (\ref{eq:phipmd2}) in the form
\begin{equation}
\phi_\pm(\vec x)={1\over2}{\sqrt{2\rho}\over m_\phi}\left(e^{\pm i\vec k_\phi\cdot\vec x}+f\,{e^{\pm ik_\phi|\vec x|}\over|\vec x|}\right) ~~ .
\end{equation}
We know from scattering theory that the term $f$ is related to the scattering cross section as $\sigma=4\pi|f|^2$. As $f$ also depends on $\sigma$, we have an algebraic equation for $\sigma$. We are interested in the behavior of $\sigma$ for small values of $k_\phi$ since for big values the interaction terms are negligible and we recover the free wave solution. At leading order in $k_\phi$ we get
\begin{equation}
\sigma \approx {4\pi\over k_\phi^2} ~~ . \label{eq:sigmad1}    
\end{equation}
This result is also shown in \cite{Jackiw:1991je}. Thus, for small values of $k_\phi$, the parameters $\varepsilon$ and $\kappa$ take the form
\begin{align}
\varepsilon &\approx 8e_mB_0R_e^2   \label{eq:epsd2}
\\
\kappa & \approx {e_mB_0R_e \over k_\phi} ~~ . \label{eq:deltad2}
\end{align}

It is important to notice that for very small values of $k_\phi$ the $1/m_\phi$ behavior of $\phi_\pm$ cancels out. From Eq. (\ref{eq:epsd2}) and Eq. (\ref{eq:deltad2}) we find that the free wave approximation used in this work is valid as long as the parameters $\varepsilon$ and $\kappa$ are much smaller than one, meaning that the electric charge and the mDM mass must satisfy
\begin{equation}
e_m \ll 2\times10^{-26}    
\end{equation}
and 
\begin{equation}
m_\phi \gg 10^{-20}\mathrm{eV}\left(e_m\over10^{-30}\right)^2 ~~ ,    
\end{equation}
respectively. These two conditions hold for the lower constraint limits shown in Fig. 2 in the main text.

\begin{figure}[ht]
    \centering
    \includegraphics[width=0.7\linewidth]{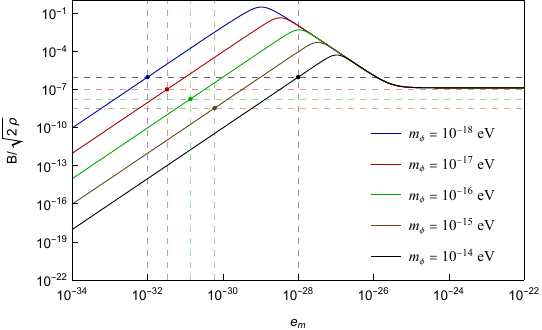}
    \caption{The dependence of $B/\sqrt{2\rho}$ on the charges of mDMs, $e_m$, for different masses, $m_\phi$.}
    \label{fig:signal}
\end{figure}

Given Eq.~(\ref{eq:phipm0d2}), we compute the magnetic field signal over the full parameter space 
of interest. The results are shown in Fig.~\ref{fig:signal}, where the normalized signal 
$B/\sqrt{2\rho}$ is plotted as a function of the millicharge $e_m$ for various $m_\phi$ values. 
As seen in the figure, $B/\sqrt{2\rho}$ stays below unity across all masses considered, indicating 
that back-reaction effects are negligible. Furthermore, the curves converge to the same 
charge-independent saturation value of $v_\phi^2\sqrt{\rho}/(2B_0) = 1.31\times10^{-7}$, 
which corresponds to a constant magnetic field signal of $1.43\,\mathrm{pT}$. Vertical grid lines 
in the plot represent the lower bounds on $e_m$ derived in this work, while horizontal grid lines 
indicate the corresponding signal strengths, i.e., the minimum detectable levels. Notably, 
all lower bounds lie in the $\varepsilon, \kappa \ll 1$ regime, validating the use of the 
free-wave approximation. However, for $e_m$ values above these lower bounds, the signal does not 
always exceed the minimum detectable value, as demonstrated in the $m_\phi = 10^{-18}\,\mathrm{eV}$ 
and $m_\phi = 10^{-14}\,\mathrm{eV}$ cases shown in Fig.~\ref{fig:signal}. This implies that, for 
some masses, an upper bound on $e_m$ also exists, as illustrated in Fig.~2 of the main text. 
In contrast, for the other masses shown in Fig.~\ref{fig:signal}, the signal always remains above 
the minimum detectable level once the lower constraint is crossed, so the entire parameter space 
above the lower limit is excluded—this applies to most of the parameter space explored in this work.

\section{Formal solution to the vector potential}\label{SuppM_1}

In this section, we show explicitly how to extract the vector potential for a given magnetic field model. The vector potential $\Aext$ can be found by solving $\vec\nabla\times\Aext=\Bext$. We write $\Bext$ and $\Aext$ in a vector spherical harmonics (VSH) expansion as
\begin{align}
\Bext(\vec x)=\sum_{\ell,m}\left(\Bextr(r)\vec Y_{\ell m}(\theta,\varphi)+\Bextps(r)\vec\Psi_{\ell m}(\theta,\varphi)+\Bextph(r)\vec\Phi_{\ell m}(\theta,\varphi)\right),
\end{align}
\begin{align}
\Aext(\vec x)=\sum_{\ell,m}\left(\Aextr(r)\vec Y_{\ell m}(\theta,\varphi)+\Aextps(r)\vec\Psi_{\ell m}(\theta,\varphi)+\Aextph(r)\vec\Phi_{\ell m}(\theta,\varphi)\right),
\end{align}
Where the ${\cal B}_i$ and ${\cal A}_i$ coefficients are also functions of $\ell$ and $m$. The curl of the background vector potential is given by 
\begin{align}
\vec\nabla\times\Aext=\sum_{\ell,m}\left(-{\ell(\ell+1)\over r}\Aextph\,\vec Y_{\ell m}-\left({d\Aextph\over dr}+{\Aextph\over r}\right)\vec\Psi_{\ell m}+\left(-{\Aextr\over r}+{d\Aextps\over dr}+{\Aextps\over r}\right)\vec \Phi_{\ell m}\right).
\end{align}
By comparison we find the following set of first order ordinary equations for the components of the vector potential
\begin{align}
    &-{\ell(\ell+1)\over r}\Aextph 
    =\Bextr, 
    \nonumber\\%\label{eq:Acurl1}
    &{d\Aextph\over dr}+{\Aextph\over r}
    =\Bextps, 
    \nonumber\\%\label{eq:Acurl2}
    &{d\Aextps\over dr}+{\Aextps\over r}-{\Aextr\over r}=\Bextph, \nonumber\\ %\label{eq:Acurl}
    &{d\Aextr\over dr}+{2\Aextr\over r}-\ell(\ell+1){\Aextps\over r}=0. \label{eq:Acurl&gauge}
\end{align}
where the last line is the gauge constraint $\vec\nabla\cdot\Aext=0$. From the first line of the Eq. (\ref{eq:Acurl&gauge}) we extract
\begin{align}
\Aextph=-\frac{r}{\ell(\ell+1)}\Bextr, \label{eq:Alm2}
\end{align}
and the second line is just a redundancy of this (from $\nabla\cdot \Bext=0$). $\Aextps$ and  $\Aextr$ can be figured out by solving the system composed by the last two lines in Eqs. (\ref{eq:Acurl&gauge}). We decouple this system finding a second order equation for $\Aextps$ given by 
\begin{align}
    \frac{\left(r^2(r\Aextps)'\right)'}{r^3}-\frac{\ell(\ell+1)}{r^2} \Aextps=\frac{1}{r^3}\left(r^3 \Bextph\right)'. \label{eq:Alm1_PDE}
\end{align} 
A formal solution for the above equation is found through
\begin{align}
  \Aextps=\int dr' G_{\ell}(r,r') \frac{1}{r'^2}{d\over dr'}\left(r'^3 \Bextph(r')\right),\label{eq:Alm1}
\end{align}
where $G_{\ell}(r,r')$ is the Green's function given by
\begin{align}
G_{\ell}(r,r') =-\frac{1}{2\ell+1}\begin{cases}
    \left(r/r'\right)^{\ell-1},\quad r<r'\\
    \left(r'/r\right)^{\ell+2}, \quad r>r'
\end{cases}.
\end{align}
Once we get $\Aextps$ explicitly, $\Aextr$ is directly found by
\begin{align}
   \Aextr=r{d\Aextps\over dr}+{\Aextps}-r \Bextph.\label{eq:Almr}
\end{align}
%From the eqs. (\ref{eq:Alm2}), (\ref{eq:Alm1}) and (\ref{eq:Almr}), one can find the $\Aext$ with given magnetic field model. 

\section{Model for the geomagnetic field and vector potential profile}

For clarity and later convenience, we note that there are three main regions for the structure of the Earth interior; the mantle, the outer core, and the inner core. These regions are separated by the Core-Mantle Boundary(CMB) and the Inner-Core Boundary (ICB), located at $R_{\mathrm{cmb}}=3486$ km and $R_{\mathrm{icb}}=1216$ km from the Earth center, respectively. The usual term `the core of the earth' refers to the combination of the inner core and outer core. We take the radius of the Earth as $R_e=6371.2$ km.

We are interested in the profile of $\Aext$ for $r\geq R_e$. From Eq. (\ref{eq:Alm2}) we get that the component $\Aextph$ is determined directly from $\Bextr$. For this we can make use of the conventional IGRF model for the Earth's geomagnetic field, which is valid for $r\geq R_\text{cmb}$ and can be written as
\begin{align}
\Bext(\vec x)=\sum_{\ell,m}C_{\ell m}\left(\frac{R_e}{r}\right)^{\ell+2}\left((\ell+1)\vec Y_{\ell m}(\theta,\varphi)-\vec\Psi_{\ell m}(\theta,\varphi)\right), \;  R_{\mathrm{cmb}}\leq r, \label{eq:B0model}
\end{align}
where the $C_{\ell m}$ are coefficients given, for instance, in reference \cite{alken2021international}. The spherical coordinates $(r,\theta,\varphi)$ are attached at the Earth center with $\theta$ and $\varphi$ being the latitude, and longitude, respectively. We have 
\begin{align}
    \Aextph=-\sum_{\ell,m}{C_{\ell m}\over\ell}{R_e^{\ell+2}\over r^{\ell+1}}, \; R_{\mathrm{cmb}} \leq r.
\end{align}

As seen from Eq. (\ref{eq:Alm1}) and Eq. (\ref{eq:Almr}), to get the profiles for the components $\Aextps$ and $\Aextr$, we need a model for the magnetic field component $\Bextph$ for all values of $r$. It is clear from the IGRF model (Eq. (\ref{eq:B0model})) that this component is absent for $r\geq R_\text{cmb}$. In fact, as will shown later, it only exists in the outer core, for $R_\text{icb}\leq r\leq R_\text{cmb}$. Unfortunately, there is no precise model for the $\Bextph$ profile. However, so far it is possible to infer a rms value of the total magnetic field over the outer core volume $V_{\rm{d}}$, defined as
\begin{equation}
{\cal B}^\text{rms}=\sqrt{\frac{1}{{V_{\mathrm{d}}}}{\int[\Bext(r,\theta,\varphi)]^2dV}}, \label{eq:Brms1}
\end{equation}
where the integration in the radial coordinate is performed in the range $R_\text{icb}\leq r\leq R_\text{cmb}$. As pointed out in Refs. \cite{2011AGUFMGP23B..01B,1995GApFD..79....1B,2010Natur.465...74G,2022NRvEE...3..255L}, the real value of ${\cal B}^\text{rms}$ ranges from 2.5 to $10\,\text{mT}$. As we only have values for the total rms magnetic field, we need to have an idea of the other components, namely, $\Bextr$ and $\Bextps$, to deduce the profile of $\Bextph$.

To construct our model for the magnetic field in the core we use the fact that it is produced by electric currents $\vec J$ that are only present in the outer core, i.e., for $R_\text{icb}\leq r\leq R_\text{cmb}$. Thus, we assume the following constraints:

\begin{itemize}
    \item [1)] the electric current merely lies in the outer core and is responsible for generating the full geomagnetic field inside and outside the Earth. The profile of the current is continuous and smooth.
    \item [2)]the inner core has a constant magnetic field with magnitude ${\cal B}_\text{ic}=6\,\text{mT}$ \cite{2024PhyS...99b5006P}.
    \item [3)] the outer core has a rms magnetic field value between $2.5\,\text{mT}$ and $10\,\text{mT}$. %$B({R_{\mathrm{icb}}})=6$ mT and $B(R_{\mathrm{cmb}})=0.42$ mT and with the constraint that  
    \item [4)] the profile for the magnetic field is continuous and smooth.
\end{itemize}

\begin{figure*}[t]
\centering
\includegraphics[width=0.6\linewidth]{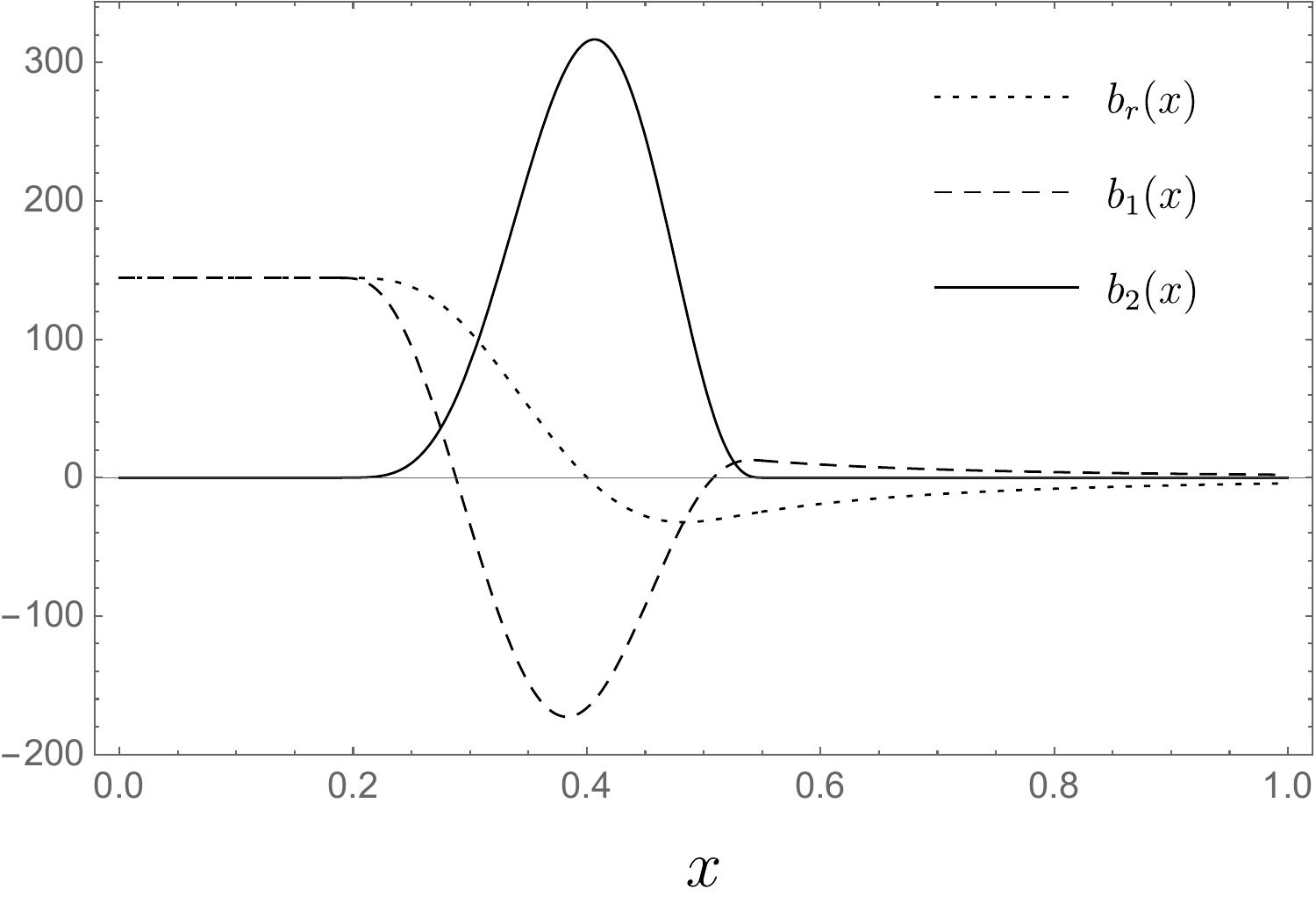}
\caption{The profiles for $b_r(x)$, $b_1(x)$, and $b_2(x)$ of the magnetic fields in the Earth interior, with $x=r/R_e$.}
\centering
\label{fig:mag_profile}
\end{figure*}

To find the magnetic field sourced by electric currents $\vec J$ in the outer core, we need to solve the system composed by the equations $\vec\nabla\times\Bext=\vec J$ and $\vec\nabla\cdot\Bext=0$. As for $\Bext$ and $\Aext$, we also write $\vec J$ in a VSH expansion as
\begin{align}
\vec J(\vec x)=\sum_{\ell,m}\left(J_r(r)\vec Y_{\ell m}(\theta,\varphi)+J_1(r)\vec\Psi_{\ell m}(\theta,\varphi)+J_2(r)\vec\Phi_{\ell m}(\theta,\varphi)\right).
\end{align}
Due to the similarities of this system to the equations used to find the vector potential, we write the same solutions
\begin{align}
 \Bextph=-\frac{r}{\ell(\ell+1)}J_r, \nonumber\\ \label{eq:Blm2}\\
 \Bextps=\int dr' G_{\ell}(r,r') \frac{1}{r'^3}{d\over dr'}\left(r'^3 J_2(r')\right),\nonumber\\ \label{eq:Blm1}\\
\Bextr=r{d\Bextps\over dr}+{\Bextps}-r J_2.\label{eq:Blmr}
\end{align}

\begin{figure*}[ht]
\centering
\includegraphics[width=0.6\linewidth]{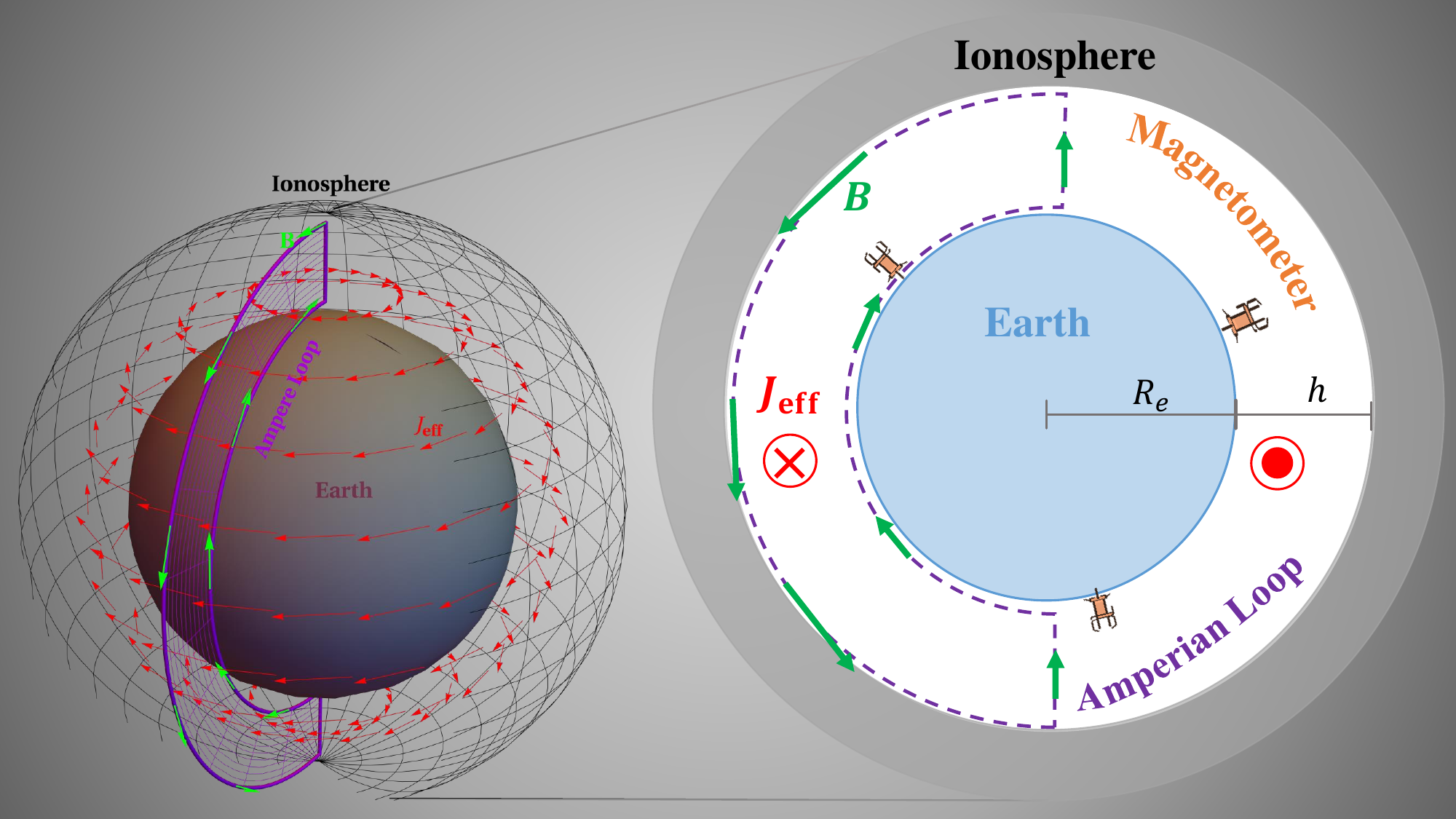}
\caption{\textbf{Right}: Natural Earth's cavity formed between the  Earth's surface and the ionosphere. It shows the tangential component of the dark matter effective current $\vec{J}_{\mathrm{eff}}$ passing through a chosen Amperian Loop. The generated magnetic field $\vec{B}$ can be probed by the magnetometers (orange symbols) placed over the surface of the Earth. This shows schematically how for tangential components the generated magnetic field signal is suppressed by the ionosphere height $h$. \textbf{Left}: 3D version of the Right.}
\label{setup-2}
\end{figure*}

From now, for convenience, we use the defined quantities $x=\frac{r}{R_e}$, $\vec b=\frac{\Bext}{B_0}$ and $\vec j=\frac{R_e\vec J}{B_0}$, where $B_0=2.94\times10^{-2}\,\text{mT}$ and is related to the coefficient $C_{10}$ in Eq. (\ref{eq:B0model}) by $C_{10}=-\sqrt{4\pi/3}B_0$. $B_0$ can also be interpreted as the value of the geomagnetic field at the equator of the  Earth surface in a simple dipole model. As the uncertainty in the values of $\cal B^\text{rms}$ is around 60\%, we only consider the coefficients $\ell=1$ in all VSH expansions since higher orders contribute with corrections of only about 10\% (at least for the IGRF model). We model the current density as the polynomial expansions
\begin{align}
    j_2 &= (a_0+ a_1x+ a_2x^2+ a_3x^3+ a_4x^4+ a_5x^5)\delta_{m0},\\
    j_r &= -2 \tilde \beta (x -x_{\rm icb})^4(x_{\rm cmb}-x)^3\delta_{m0}. 
\end{align}
From $\vec\nabla\cdot\vec j=0$ it is found that $j_1=j_r+xj_r'/2$. The coefficients $a_i$ are found after imposing the boundary conditions
\begin{align}
& j_2(x_{\rm icb})=0, \quad j_2(x_{\rm cmb})=0,\nonumber\\
& j_2'(x_{\rm icb})=0, \quad j_2'(x_{\rm cmb})=0,\nonumber\\
& b_1(x_{\rm icb})=\frac{1}{\sqrt{2}}\frac{{\cal B}_{\rm ic}}{{\rm C_{10}}}, \quad b_1(x_{\rm cmb})= \sqrt{\frac{4\pi}{3}}\frac{1}{x_{\rm cmb}^3},
\end{align}
where $b_1$ is calculated through Eq. (\ref{eq:Blm1}). The profile of $j_r$ satisfies a vanishing value for $j_r$, $j_1$ and their derivatives at the outer core boundaries. The $(x-x_{\rm icb})^4$ of the $j_r$ profile makes the current density tends to be more prominent at locations closer to the boundary with the mantle. Finally, we use Eq. (\ref{eq:Brms1}) to calculate $\tilde\beta$, assuming the value for ${\cal B}^\text{rms}$ ranges between $2.5$ and $10\,\text{mT}$. We get $\tilde\beta=1.2\times10^8\beta$, where $\beta$ is a value ranging from 1 to 4.6, that accounts for the uncertainty in ${\cal B}^\text{rms}$. The profiles for the three components of the magnetic fields in the Earth interior are plotted in Fig. \ref{fig:mag_profile}.

%The choices $m=0$ for $j_2$ and $m=1,-1$ for $j_r$ make sense with angular distribution and

Using Eq. (\ref{eq:Alm1}) and Eq. (\ref{eq:Almr}) and integrating by parts, we write the final expression for the vector potential as
\begin{align}
\Aext = -B_0 R_e\left(1.01\beta\left(R_e\over r\right)^3(2\vec Y_{10}-\vec\Psi_{10})-\sqrt{4\pi\over3}\left(R_e\over r\right)^2\vec\Phi_{10}\right), \quad R_e<r. \label{eq:A0solsupp}
\end{align}
As a final note, in Fig. \ref{setup-2}, we show the illustration for the subdominant signal from the tangential effective current $\vec{J}_\text{eff}=J_\text{eff,t}\vec\Phi_{10}$ that is proportional to the second term in Eq. (\ref{eq:A0solsupp}). From the Ampere's law $\oint d\vec l\cdot\vec B=\int d\vec S\cdot\vec J_\text{eff}$, the 
Ampere loop showed in the Fig. \ref{setup-2} gives us $2\pi R_e B \approx 2\pi R_eh J_\text{eff,t}$ and thus $B \approx R_e(h/R_e) J_\text{eff,t}$ suppressed by the height of the ionosphere $h\ll R_e$. While the radial effective current $\vec{J}_\text{eff}=J_\text{eff,r}\vec Y_{10}$ give us $2\pi R_e B \approx \pi R_e^2 J_\text{eff,r}$ and thus is dominant signal of the form $B \approx R_e J_\text{eff,r}/2$ as shown in the Fig. 1 of the main text.

\end{document}